\documentclass[twocolumn,aps,prl,superscriptaddress]{revtex4}
\usepackage{amssymb}
\usepackage{amsmath}
\usepackage{color}
\usepackage{graphicx}
\usepackage{marvosym}
\usepackage{ulem}
\usepackage{notoccite}

\usepackage{comment}

\usepackage{bbm}

\usepackage{graphicx,tabularx}
\usepackage{dcolumn}
\usepackage{bm}
\usepackage{xcolor}

\usepackage{marvosym}

\newcommand{\be}{\begin{equation}}
\newcommand{\ee}{\end{equation}}

\newcommand{\la}{\langle}
\newcommand{\ra}{\rangle}

\newcommand{\Tr}{{\rm \, Tr\,}}

\newcommand{\ket}[1]{| #1\ra}
\newcommand{\avg}[1]{\langle #1 \rangle}

\renewcommand{\vec}[1]{{\bf #1}}

\begin{document}

\title{Multistable excitonic Stark effect}

\author{Ying Xiong}
\affiliation{Division of Physics and Applied Physics, Nanyang Technological University, Singapore 637371}
\author{Mark S. Rudner}
\affiliation{Center for Quantum Devices and Niels Bohr International Academy, Niels Bohr Institute, University of Copenhagen, 2100 Copenhagen, Denmark}
\author{Justin C.W. Song}
\affiliation{Division of Physics and Applied Physics, Nanyang Technological University, Singapore 637371}
\affiliation{Institute of High Performance Computing, A*STAR, Singapore, 138632}

\begin{abstract}
  The optical Stark effect is a tell-tale signature of coherent light-matter interaction in excitonic systems, wherein an irradiating light beam tunes exciton transition frequencies.
Here we show that, when excitons are placed in a nanophotonic cavity, the excitonic Stark effect can become highly nonlinear, exhibiting
multi-valued and hysteretic Stark shifts that depend on the history of the irradiating light.
This multistable Stark effect (MSE) arises from feedback between the cavity mode occupation and excitonic population, mediated by the Stark-induced mutual tuning of the cavity and excitonic resonances.
Strikingly, the MSE manifests even for very dilute exciton concentrations and can yield 
discontinuous Stark shift jumps of order meV.
We expect that the MSE can 
be realized in readily available transition metal dichalcogenide excitonic systems placed in planar photonic cavities, at modest pump intensities.
This phenomenon can provide new means to engineer coupled 
states of light and matter that can persist even in the single exciton limit.  
\end{abstract}

\maketitle

Strong light-matter interaction can provide a 
versatile platform for dynamically controlling quantum matter~\cite{basov}.
A striking example is the excitonic optical Stark effect~\cite{stark_morkoc, stark_chemla, stark_townes, stark_GaAs}: 
off-resonant irradiation of an excitonic system, with frequency below the exciton transition energy, continuously blue-shifts the exciton transition to higher frequencies as the light intensity increases~\cite{stark_townes, stark_GaAs, stark, Combescot, Hayat}. 
In contrast to the fixed Rabi splitting found for polaritons, that is independent of the intensity of light~\cite{polariton_hopfield,polariton_weisbuch,TMD_polariton}, the optical Stark effect is linear in the irradiation intensity. This dependence grants on-demand tunability of excitonic properties.
In transition metal dichalcogenides (TMDs), Stark shifts are furthermore sensitive to light polarization, thereby enabling direct control over the valley excitons necessary for valley opto-electronics~\cite{stark, stark_BS, stark_biexciton}. 

\begin{figure} [t]
    \centering
    \includegraphics[width=\columnwidth]{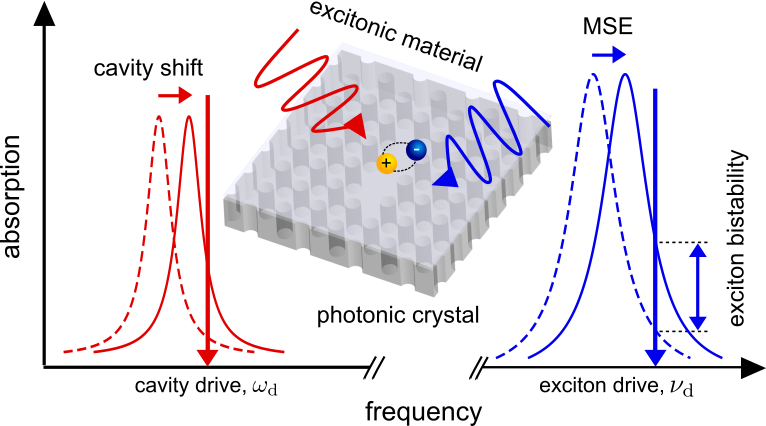}
      \caption{ Mutual tuning of exciton (right panel) and cavity (left panel) transitions induced by the optical Stark effect, wherein the exciton transition frequency is sensitive to the cavity mode occupation (and vice versa).
When the excitons and the cavity mode are simultaneously pumped
(downward arrows indicate the corresponding pump frequencies),  
the exciton and cavity transitions can shift into resonance with their drives (from dashed to solid curves).
These population-induced shifts generate the feedback loop that gives rise to the 
multistable Stark effect (MSE).  (inset) A two-dimensional excitonic material such as a transition metal dichalcogenide 
can be readily layered on top of a planar nanophotonic cavity formed by a photonic crystal defect to achieve the conditions for realizing the MSE.}
\label{schematic}
\end{figure}

Here we propose that the optical Stark effect can take on a markedly different character when an excitonic system is placed in a nanophotonic cavity (Fig.~\ref{schematic} inset).
In this setting, the Stark shift becomes a dynamical variable, 
with the cavity field taking on the role of the irradiating field that shifts the excitonic levels.
In particular, when the excitonic and cavity modes are simultaneously pumped, 
the optical Stark effect can become {multistable}, exhibiting a hysteretic Stark shift that depends on the history of the optical drive.
As we explain below and indicate in Fig.~\ref{schematic}, this multistable Stark effect (MSE) arises due to a Stark-induced mutual tuning: the excitonic transition frequency (right panel)
is sensitive to the cavity mode occupation, while the cavity resonance (left panel) 
is sensitive to the exciton population. 
When applied 
exciton and cavity driving fields are detuned from their respective bare transition frequencies, the mutual tuning sets up a feedback that shifts the exciton and cavity transition frequencies into resonance with their respective drives (dashed to solid lines, Fig.~\ref{schematic}).
This feedback leads to a highly non-linear, and multistable Stark effect.  

The MSE features discontinuous transitions between multiple distinct steady states of the combined cavity-exciton system, and can exhibit large discontinuous Stark shift jumps of order meV. Indeed, we find that the exciton population can take on multiple steady state values (Fig. 2a) with a hysteretic behavior that is controlled by a weak
cavity drive far-detuned from the original exciton resonance.
Further, the magnitude of the Stark shift jump from one stable state to another can be directly tuned by 
the drive that pumps the excitonic population. These mechanisms provide in-situ means of tailoring the switching behavior in the exciton/cavity system.

We expect that the MSE can be realized in 
TMDs (e.g., WS$_2$) on currently available high quality factor planar photonic cavities ~\cite{PhC1, PhC2, PhC3, PhC4} (Fig.~\ref{schematic} inset), 
even at low optical drive strengths of several to tens of ${\rm kW}/{\rm cm}^2$. 
This platform 
 provides new means of constructing 
hysteretic, nonlinearly coupled states 
of light and matter that can, in principle, persist even at the single exciton limit.

{\it Stark-induced mutual tuning and nonlinearity.---}  
The key to achieving the MSE is the nonlinearity mediated by strong coupling between cavity photon modes and excitons.
As we now explain, this nonlinearity in the cavity-exciton system can arise directly through the Stark effect. 
As a simple and clear illustration of the MSE, we first focus on 
a single localized excitonic mode 
interacting with a single cavity mode (of a single polarization). 
We will discuss 
the MSE for delocalized excitons in an extended 2D excitonic layer later in the text. 

We model the localized exciton mode as a simple two-level system with 
bare resonance angular frequency $\nu^{(0)}$; we denote the ground state (no exciton) by $\ket{P = 0}$, 
  and the excited state (exciton present) by $\ket{P = 1}$. 
The cavity photon mode has angular frequency $\omega^{(0)}$.
In the dispersive limit, the dynamics of the system are described by the Hamiltonian $H = H_X + H_0 + H_{\rm int}$ with (setting $\hbar = 1$ here and throughout, unless otherwise stated):
\begin{eqnarray}
\label{H one-valley}
H_X = \nu^{(0)}  \hat{P}, \ \ H_0 =   \omega^{(0)} a^\dagger a, \ \ H_{\rm int} = V a^\dagger a \hat{P},
\end{eqnarray}
where $a^\dagger$ is the creation operator for the cavity photon mode, and $\hat{P} = s^z + \mathbbm{1} /2$ counts the exciton population via $\hat{P}\ket{P} = P \ket{P}$, where $\mathbbm{1}$ is the $2 \times 2$ identity matrix and $s^z = \sigma^z/2$, where $\sigma^z$ is the third Pauli matrix.

The last term in Eq.~(\ref{H one-valley})  
encodes a dispersive coupling, $V$, between the excitons and cavity photons, that is valid for $V \ll |\nu^{(0)} - \omega^{(0)}|, \omega^{(0)}, \nu^{(0)}$.
In this limit, the magnitude of $V$ can be controlled directly through engineering of the microcavity mode profile
and its detuning from the exciton resonance, $\Delta = \nu^{(0)} - \omega^{(0)}$.
Throughout this work we will consider $\Delta >0 $, which ensures that $V>0$;
for a derivation of the dispersive coupling $V$ and parameter estimates for 
a TMD/cavity system, see Supplementary Information ({\bf SI}). 

Crucially, through the dispersive coupling, both the exciton and cavity photon 
resonances are mutually dependent on the other's occupation.
For a state with $m$ cavity photons present, and excitonic state $P = \{0, 1\}$, the (cavity-dressed) exciton and (exciton-dressed) cavity photon resonance angular frequencies, $\tilde{\nu}(m)$ and $\tilde{\omega}(P)$, respectively, are given by: 
\be
\label{eq:mutualtuning}
\tilde \nu(m) = \nu^{(0)} + Vm, \quad \tilde \omega (P) = \omega^{(0)} +  V P. 
\ee
The excitonic resonance $\tilde \nu(m)$ experiences a blue shift away from its bare resonance frequency that is proportional to the photon number in the cavity -- the optical Stark effect~\cite{Combescot, stark,stark_townes,stark_chemla}.
We characterize this by the excitonic Stark shift: $\delta E \equiv \tilde \nu(m)  -  \nu^{(0)} = Vm$.
Similarly, the 
cavity photon resonance frequency 
$\tilde \omega (P)$ also depends on the occupation of the excitonic state, shifting as $P$ changes.
The mutual tuning of exciton and cavity photon transitions exhibited in Eq.~(\ref{eq:mutualtuning}) provides a natural means of feedback, and as we now discuss, 
gives rise to nonlinear dynamical phenomena in the system. 

{\it Multistable Stark effect and cavity-exciton steady states.---} To demonstrate the MSE, we consider an exciton-photon microcavity system with 
laser drives at angular frequencies $\nu_d$ and $\omega_d$.
These fields 
pump the excitonic and cavity photon modes, 
respectively.
This selectivity can be achieved by choosing $\nu_d$ and $\omega_d$ to be 
slightly detuned from $\tilde \nu (0)$ and $\tilde \omega (0)$, respectively, with their individual detunings much smaller than $\Delta$.

In the presence of these laser driving fields, the 
Hamiltonian becomes $\mathcal{H}(t) = H + H^{(d)}_{X}(t)+ H^{(d)}_{0}(t)$, with
\begin{eqnarray} \label{eq:driven H}
H_X^{(d)}(t) &=& \frac{F_X}{2} \big(e^{-i\nu_d t} \vec \sigma^+ + h.c.), \nonumber \\ 
 H_0^{(d)}(t) &=& \frac{F_0}{2} (e^{-i\omega_d t} a^\dagger + h.c.), 
\end{eqnarray} 
where $F_X$ and $F_0$ are the drive amplitudes, and $\sigma^+ = (\sigma^x + i \sigma^y)/2$, where $\sigma^{x,y}$ are the $x,y$ Pauli matrices.
In anticipation of 
making a rotating wave approximation below, 
  we have discarded counter-rotating terms in Eq.~(\ref{eq:driven H}). 
  
\begin{figure} [t]
    \centering
    \includegraphics[width=\columnwidth]{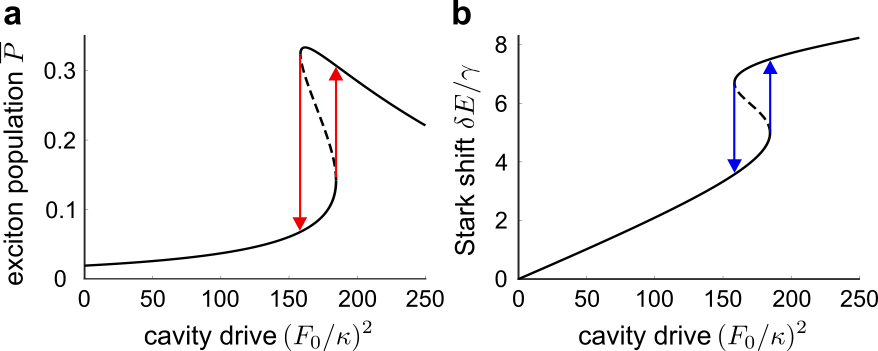}
    \caption{A single excitonic emitter coupled to a cavity can display 
      bistable and hysteretic steady-states of the (a) exciton and (b) cavity photon populations (reflected in the Stark shift, $\delta E$).
      The steady states are obtained 
      by solving Eqs.~(\ref{eq:steadys}) and (\ref{eq:steadym}); 
    the thick solid lines indicate the stable solutions, and the thin dashed lines indicate the unstable solutions, see {\bf SI}. Illustrative dimensionless parameters used: $F_X/\gamma = {2}$, $V/\gamma = {0.25}$, $\nu^d - \nu (0) ={7} \gamma$,  $\omega^{d} - \tilde \omega(0) ={1.5} \kappa$ and $\gamma/\kappa = 10$.   }
    \label{fig2}
\end{figure}

To explicitly demonstrate the MSE, we track the exciton and cavity photon populations in the driven system 
in the presence of Markovian dissipation that accounts for exciton relaxation (recombination) and cavity photon loss.
As a first step, we transform into a frame that co-rotates with the drives, 
using $U(t)=\exp{(-i \omega_{d}t\, a^\dagger a  - i \nu_{d}t\, \hat{P} )}$. 
In the rotating frame, the system evolves according to the (static) Hamiltonian 
$\tilde{\mathcal{H}}~=~\tilde{H}_X + \tilde{H}_0 + \tilde{H}_{\rm int}$, with 
$\tilde{H}_X =  (\nu^{(0)} - \nu_{d} ) \hat{P} + F_X\sigma^x$/2,
and 
$\tilde{H}_0 = (\omega^{(0)} - \omega_{d}) a^\dagger a + F_0(a^\dagger + a)/2$.
The interaction $\tilde{H}_{\rm int} = V a^\dagger a \hat{P}$ does not change under the transformation. 

Using the rotating-frame Hamiltonian $\tilde{\mathcal{H}}$, we take the 
 density matrix of the composite exciton and cavity system (in the rotating frame), $\tilde{\rho} (t)$, to evolve according to the master equation 
\be \label{dynamics}
\partial_t \tilde{\rho}= i [\tilde{\rho} (t), \tilde{H}_0 +  \tilde{H}_X +  \tilde{H}_{\rm int} ] + \mathcal{I}_X[\tilde{\rho} (t)] + \mathcal{I}_0[\tilde{\rho} (t)], 
\ee
where  $\mathcal{I}_X[\tilde{\rho}] = \gamma (2 \sigma^- \tilde{\rho} \sigma^+ - \sigma^+ \sigma^- \tilde{\rho} - \tilde{\rho} \sigma^+ \sigma^-)$ accounts for 
recombination of the exciton, with rate $\gamma$, and $\mathcal{I}_0[\tilde{\rho}] = \kappa (2 a \tilde{\rho} a^\dagger - a^\dagger a\tilde{\rho}  - \tilde{\rho} a^\dagger a )$ describes losses in the microcavity photon mode with rate $\kappa$.
The interaction $\tilde{H}_{\rm int} \neq 0$ couples the cavity and exciton subsystems by the mutual tuning of their transition frequencies as in Eq.~(\ref{eq:mutualtuning}).  

While Eq.~(\ref{dynamics}) can generically encode a variety of complex dynamical regimes 
of the composite system, as we now discuss, a large separation in the 
cavity and exciton decay timescales enables direct evaluation of the MSE steady states (cf.~Ref.\cite{Rudner} for general discussion). 
Indeed, the regime wherein the excitonic system relaxes far faster than the cavity photon system 
can be readily achieved in many exciton-cavity setups, see estimate below. 
Physically, this separation of timescales means that the reduced density matrix of the excitonic system $\tilde{\rho}_X (t) \equiv {\rm Tr}_0\, \tilde{\rho}(t) $ rapidly reaches a quasistationary state over a time that is short compared with 
the characteristic evolution timescale of the cavity photon; here ${\rm Tr}_{0}$ [${\rm Tr}_X$] denotes the partial trace 
over photonic [excitonic] degrees of freedom.
On the timescale of excitonic relaxation, the cavity state  
$\tilde{\rho}_0(t) \equiv {\rm Tr}_X \tilde{\rho}(t)$ 
can be treated as 
quasistatic, allowing the formation of an excitonic steady state that depends parametrically on $\tilde{\rho}_0$. 
On the timescale that the cavity state $\tilde{\rho}_0(t)$ evolves, $\tilde{\rho}_X (t)$ maintains a quasistationary state that adiabatically follows the slow evolution of $\tilde{\rho}_0(t)$.

Using this separation of timescales, in describing the 
time evolution of the exciton and cavity photons we adopt a mean-field decoupling~\cite{Rudner} of
Eq.~(\ref{dynamics}) by replacing the cavity-exciton coupling by its mean-field averages ${\rm Tr}_0(\tilde{\rho} (t) \tilde{H}_{\rm int}) \to  V \la m (t) \ra \hat{P} $ 
and ${\rm Tr}_X(\tilde{\rho} (t) \tilde{H}_{\rm int}) \to V a^\dag a \la \hat{P} (t) \ra $, where $\la \hat{P} (t) \ra \equiv {\rm Tr} [\hat P \tilde{\rho}_X (t)]$ and  $m (t) \equiv {\rm Tr} [a^\dag a \tilde{\rho}_0 (t)]$.
This mean-field decoupling is justified in the semiclassical regime  where the photon number in the cavity is large and fluctuations are small~\cite{Rudner}.
With this mean-field decoupling, the 
(rotating frame) exciton and cavity density matrices $\tilde{\rho}_X (t)$ and $\tilde{\rho}_0 (t)$, respectively, evolve according to: 
\begin{align}
\label{eq:decoupled}
\partial_t \tilde{\rho}_X (t) & = i \left[\tilde{\rho}_X (t) , \tilde{H}_X + V  \la m (t) \ra s \right] + \mathcal{I}_X[\tilde{\rho}_X (t)], \\
\label{eq:decoupled2}
\partial_t \tilde{\rho}_0 (t) & = i \left[\tilde{\rho}_0 (t) , \tilde{H}_0+ V a^\dag a \la \hat{P} (t)  \ra \right] + \mathcal{I}_0[\tilde{\rho}_0 (t)].   
\end{align}

The exciton population dynamics can be obtained by directly evaluating 
the elements of $\tilde{\rho}_X (t) $ in Eq.~(\ref{eq:decoupled}) to obtain effective Bloch equations.
Writing
$\la s^i  (t) \ra \equiv \Tr [s^i \tilde{\rho}_X (t)  ]$ where $s^i = \sigma^i/2$ for $i = x, y, z$, and noting $\Tr [\tilde{\rho}_X (t) ] = 1$, we obtain 
\begin{eqnarray}
\label{eq:sdynamics}
\partial_t\la s^x (t)\ra  &=& \delta \nu (t) \la s^y (t) \ra -  \gamma \la s^x(t) \ra, \nonumber \\
\partial_t\la s^y (t) \ra  &=& - F_X \la s^z (t) \ra  - \delta \nu (t) \la s^x (t) \ra -\gamma \la s^y (t) \ra, \nonumber \\
\partial_t\la s^z (t) \ra  &=& F_X \la s^y (t)\ra - 2\gamma (\la s^z (t)\ra + 1/2), 
\end{eqnarray}
where $\delta \nu(t) =  \nu_{d} - \tilde \nu [\la m (t) \ra]$. 
We solve for the excitonic (quasi)-steady state by setting the three equations above equal to zero, and assuming that the cavity 
mode occupation $\avg{m (t) } = m$ is fixed. 
We thus obtain the (quasi)-steady-state population of the excitonic mode as a function of the cavity occupation, $m$:  
\be
\label{eq:steadys}
\overline{P} (m) =  \overline{\la s^z\ra} + \frac{1}{2} = \frac{F_X^2 /2}{F_X^2+ 2\left[ \gamma^2 +\big(\nu_{\rm d} - \tilde{\nu}(m)\big)^2 \right]}, 
\ee
where $\overline{\la s^z\ra}$ is the time independent steady state solution of Eq.~(\ref{eq:sdynamics}). As evident from Eq.~(\ref{eq:steadys}), the steady state excitonic population depends both on the excitonic drive strength, $F_{\rm X}$, 
and parametrically on the cavity population through the stark-shifted exciton resonance, 
$\tilde{\nu}(m)$. 

The steady state cavity population 
can be obtained heuristically by first considering the familiar expression 
for the average population of a 
driven cavity mode 
with a fixed resonance frequency, $\omega$: 
  $\bar{m} = (F_0^2/4)/\{\kappa^2 + (\omega_{\rm d} - \omega)^2\}$.
  Due to the Stark-induced mutual tuning described above, the cavity resonance frequency changes with the exciton population, see Eq.~(\ref{eq:mutualtuning}). As a result, we replace $\omega \to  \tilde \omega [\overline{P}(m=\bar{m})]$ to yield a self-consistency relation for the cavity mode population: 
\be
\label{eq:steadym}
\bar{m} = (F_0^2/4)/\{\kappa^2 + \big(\omega_{\rm d} - \tilde \omega [\overline{P} (\bar{m}) ]\big)^2\}.
\ee
We note that this heuristically-obtained self-consistency relation agrees with 
results obtained through careful analysis of the evolution of the full density matrix of the joint system~\footnote{For discussions of non-adiabatic effects and switching near bifurcation points, see Ref.~\cite{Rudner}}, in the regime $\kappa/\gamma \ll 1$, and $V^2/\gamma \ll \kappa$~\cite{Rudner}. 
The steady state cavity occupation thus depends on the 
steady state exciton population through its mutually-tuned cavity transition in Eq.~(\ref{eq:mutualtuning}). 

We now explicitly exhibit the multistability described by Eqs.~(\ref{eq:steadys}) and (\ref{eq:steadym}). Choosing drive frequencies slightly blue detuned from the bare exciton and cavity resonances (see Fig.~\ref{fig2} and caption for parameter values), Eqs.~(\ref{eq:steadys}) and (\ref{eq:steadym})
  yield multiple solutions for $\overline{P}$ 
  as a function of $F_0$ 
  (for all other parameters held fixed in this regime). These multiple steady states arise from the MSE, as evidenced by 
  the jumps of the Stark shift $\delta E$ (on the order of the exciton decay rate $\gamma$) displayed in Fig.~\ref{fig2}b. 

  Within the bistable regime, two distinct stable steady-state solutions for $\overline{P}$ and $\delta E$ 
  exist for the same drive parameters (solid lines). This enables a hysteretic behavior of the excitonic system that depends on the history of the optical drive. Indeed, as $F_0$ increases from zero (forward sweep), $\overline{P}$ (as well as $\bar{m}$) 
  jump to the upper branch of solutions (upward arrow) at a forward threshold amplitude. However, when $F_0$ is then decreased (reverse sweep), both  $\overline{P}$ (and $\delta E$) jump to the lower branch of solutions (downward arrow) at a distinct reverse threshold amplitude. This hysteresis enables the system to operate as an optically-controlled ``exciton switch'' with ``off'' and ``on'' states as the lower and upper branches. Strikingly, this excitonic hysteresis occurs even for a single excitonic mode, in sharp contrast to other nonlinearities induced at high exciton density~\cite{polariton_bistability}.

  {\it MSE in transition metal dichalcogenides.---} Having exhibited the MSE mechanism for a single excitonic emitter, we now discuss the MSE in readily available two-dimensional excitonic systems. A natural class of candidate materials are the atomically thin 
 TMDs, which possess room temperature stable excitons and large Stark effects~\cite{stark, stark_BS, stark_biexciton}, and can be easily 
  integrated with planar photonic crystal cavities, as in the inset of Fig.~\ref{schematic}.
Here we will focus on zero center of mass momentum (COMM) excitons in a single valley, 
where excitons obey circular polarization selection rules~\cite{TMD_exciton_review, TMD1, TMD2, TMD3, TMD_valley, intervalley_scatter}: by driving the TMD with circularly polarized light of fixed handedness with frequency close to the exciton resonance, only excitons in the corresponding valley will be excited. 

To describe the MSE in TMDs, we consider an extended TMD layer placed on top of a photonic cavity, see Fig.~\ref{schematic} inset.
We first note that the TMD excitonic mode at $\nu^{(0)}$ can have a large effective degeneracy $\mathcal{N}$.
This degeneracy accounts for excitons at distinct exciton center of mass spatial coordinates; these degenerate exciton emitters can form plane wave superpositions that lead to delocalized excitonic modes~\cite{Rana,Lee,Yamamoto}. 
Importantly, the modes with zero COMM interact coherently (in phase) with the same cavity photonic mode~\cite{Rana,Lee,Yamamoto} (with a wavelength of a few hundred nanometers); similarly, for exciton pumping fields 
that have large wavelengths of order several hundred nanometers, multiple excitonic emitters can be driven in phase with each other.
As such, in describing the TMD layer excitonic-cavity system, we replace $\hat{P} \to \hat{P}_{\rm tot} = \sum_{j} \hat{P}_j$ in the Hamiltonian Eq.~(\ref{H one-valley}), as well as $\sigma^{+,-} \to s^{+,-}_{\rm tot} = \sum_{j} \sigma_{j}^{+,-}$ in Eq.~(\ref{eq:driven H}), where the sum over $j$ runs over each of the $j=1, \ldots, \mathcal{N}$ degenerate excitonic emitters. Similarly, $\tilde{\omega} (P_{\rm tot}) \to \tilde{\omega} = \omega^{(0)} + V P_{\rm tot}$ in Eq.~(\ref{eq:mutualtuning}) where $P_{\rm tot} = 0 , 1, 2,  \cdots $ are eigenvalues of $\hat{P}_{\rm tot}$. Since all the emitters interact with the same cavity photon mode, $\tilde{\nu}(m)$ in Eq.~(\ref{eq:mutualtuning}) remains unchanged. 

We follow a similar procedure and use the separation of timescales as discussed above for tracking the exciton and cavity photon populations (see {\bf SI} for full details). In so doing, we take a spin-coherent-state ansatz so that the dynamics of the multiple emitter system can be 
analyzed in terms of the dynamics of a giant spin ${\vec s}_{\rm tot} = s^{x}_{\rm tot} \hat{\vec x} + s^{y}_{\rm tot} \hat{\vec y}+ s^{z}_{\rm tot} \hat{\vec z}$, where $s^{x,y,z}_{\rm tot} = \sum_i s^{x,y,z}_i$ is summed over the excitonic emitters. For fixed cavity occupation $m$ we obtain the steady-state exciton population in the extended system (see {\bf SI}) as 
\be
\label{eq:stot}
\overline{P}_{\rm tot} = \frac{\mathcal{N} F_X^2/2}{F_X^2+ 2(\Gamma^2 + [\nu_d- \nu(m)]^2)}, 
\ee
where $\Gamma$ is the exciton recombination rate (for the zero COMM excitons) in the extended TMD system; we note, parenthetically, that this rate can be estimated from the recombination rate $\gamma$ of a single localized exciton emitter as $\Gamma \sim \mathcal{N} \gamma$~\cite{Rana, Lee, Yamamoto}. In obtaining Eq.~(\ref{eq:stot}) we have taken a large degeneracy $\mathcal{N} \gg 1$ as well as focused on the low-excitation regime.

\begin{figure}
    \centering
    \includegraphics[width=\columnwidth]{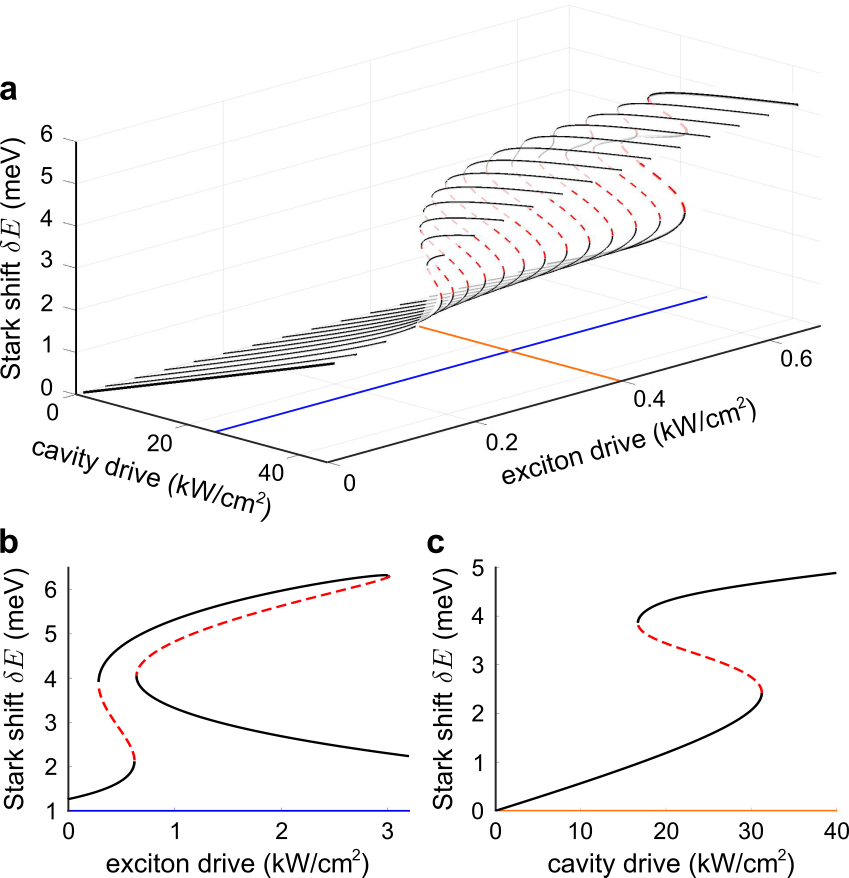}
    \caption{(a) MSE shift $\delta E$ of the excitonic system in a monolayer TMD coupled to a cavity obtained from Eq.~(\ref{eq:steadym}) and (\ref{eq:stot}) displaying multiple steady states (bistable [low exciton drive] and then tristable [larger exciton drive]) and discontinuous $\delta E$ 
    jumps. Panel (b) shows line cut of $\delta E$  
    as a function of exciton drive at a fixed cavity drive of $25$ kW/cm$^2$ [as indicated by the blue line in (a)]. Panel (c) displays  
    $\delta E$ as a function of cavity drive at a fixed exciton drive of $0.4$ kW/cm$^2$ [as indicated by the orange line in (a)]. Solid lines indicate stable solutions; whereas dashed lines indicate unstable states. Here we used parameters: $\Gamma = 1\, {\rm meV}$, $\nu_d - \nu(0) = 4\, {\rm meV}$, dispersive coupling $V= 0.2\, {\rm meV}$, $\kappa = 0.1\, {\rm meV}$, $\mathcal{N} = 1000$ and $\omega_d - \omega(0) = 0.2 \, {\rm meV}$, see {\bf SI} for detailed estimates and discussion of parameters.} 
    \label{fig3}
\end{figure}

Before we exhibit the MSE in TMD systems, we first discuss the parameters for the cavity-exciton system. We note that the excitonic mode degeneracy $\mathcal{N}$ can be large and can range from $\mathcal{N} \sim 10^2 - 10^4$~\cite{Rana}; this arises from the large number of excitonic modes that can interact coherently within a single wavelength of either the cavity photon mode or the exciton drive~\cite{Rana,Lee,Yamamoto}. An estimate of $\mathcal{N}$ can be obtained from the ratio of the mode area of the photonic mode (the square of its wavelength) and the effective size of an exciton (the square of its Bohr radius)~\cite{Yamamoto}. 
Further, recombination times for zero COMM excitons in typical monolayer TMDs can range from $\Gamma^{-1} \sim $ $0.5$ to a few picoseconds \cite{exciton_lifetime, WSe2_lifetime, exciton_lifetime_2ps, WS2_lifetime, MoS2_hBN}, whereas cavity relaxation times can be as long as tens to a hundred picoseconds \cite{PhC1, PhC2, PhC3, PhC4}. As a result, $\kappa \ll \Gamma$, justifying the separation of time scales and mean-field decoupling approach we have used to describe the MSE. Lastly, strong light-matter interaction in monolayer TMDs~\cite{stark, stark_BS} can lead to sizeable values of dispersive coupling $V \approx {0.1 - 0.5} \, {\rm meV}$, see {\bf SI} for a detailed estimate. {In the plots we have chosen $V=0.2 \, {\rm meV}$ for illustration.}

Solving Eq.~(\ref{eq:stot}) together with Eq.~(\ref{eq:steadym}) yields an excitonic multistability and MSE, as shown in Fig.~\ref{fig3}. With realistic parameters for monolayer WS$_2$ and photonic crystal cavities, discontinuous jumps in the excitonic Stark shift can be readily achieved by moderate cavity and exciton drive intensities of order kW/cm$^2$. 

Interestingly, distinct regimes of multistability can be accessed; at low exciton drive strength a bistable MSE manifests (as cavity drive is swept) whereas larger exciton drives display tristabilities (see Fig.~\ref{fig3}a,b). 
Indeed, the MSE displays hysteretic behavior as either exciton or cavity drives are swept, with  Fig.~\ref{fig3}b,c displaying sizeable discontinuous $\delta E$ of order meV. We note that together with multistable $\delta E$ shown in Fig.~\ref{fig3}, the exciton population  $\overline{P}_{\rm tot}$ similarly exhibits multistability and hysteresis (see also Fig.~\ref{fig2}).
While we have focused on the MSE and its concomitant excitonic multistability, multiple stable states of the cavity mode (so-called ``optical multistability,'' characterized by distinct steady state values of $\bar{m}$) can also arise via the MSE. (Note that in Fig.~2b, $\delta E$ is directly proportional to the cavity photon occupation.) This effect is similar to dispersive optical multistability in highly nonlinear optical media~\cite{bonifacio,hassan1978, gibbs, bowden,hermann, Ginossar2012}, and may provide new means for controlling optical states. 

From a fundamental perspective, the MSE arises from the fact that the excitonic Stark effect is 
an inescapable consequence of the partial fermionic nature of excitons~\cite{Combescot} -- a property that is present even in dilute exciton systems. 
Indeed, we find in Fig.~\ref{fig3} that a MSE manifests 
for a steady state excitonic population on the order of $\overline{P}_{\rm tot} \sim 1$, see {\bf SI}. 
This indicates that the MSE occurs even as approximately one exciton is excited in the entire photonic cavity (corresponding to a low exciton density of order $10^{10}\ {\rm cm}^{-2}$). This distinguishes MSE and its associated nonlinear phenomena from other types of multistable behavior, e.g., optical bistabilities that originate from exciton-exciton interactions that typically require a large density of excitons to enable bistable behavior~\cite{polariton_bistability, bistability_phototherm1}.
Perhaps most exciting is how MSE-induced hysteresis in the exciton population as a function of optical drive yields jumps in $\overline{P}_{\rm tot}$ of order unity; these may provide controllable means of selectively exciting/de-exciting a single exciton as well as controlling its emission.

\begin{acknowledgments}
J.C.W.S. acknowledges support from the National Research Foundation (NRF), Singapore under its NRF fellowship programme award number NRF-NRFF2016-05, the Ministry of Education, Singapore under its MOE AcRF Tier 3 Award MOE2018-T3-1-002, and a Nanyang Technological University start-up grant (NTU-SUG). M.R. gratefully acknowledges the support of the Villum Foundation, and the
European Research Council (ERC) under the European Union Horizon 2020 Research and Innovation Programme (Grant Agreement No. 678862).
\end{acknowledgments}

\clearpage

\newpage

\setcounter{equation}{0}
\setcounter{figure}{0}
\renewcommand{\theequation}{S\arabic{equation}}
\renewcommand{\thefigure}{S\arabic{figure}}

\onecolumngrid

\section{Supplementary Information for ``Multistable excitonic Stark effect''}

\subsection{Optical Stark effect}

For the convenience of the reader, in this section we review the optical Stark effect and how it arises from the light-matter interaction between an exciton and light. In so doing, we concentrate on a single localized exciton emitter and its interaction with a single cavity mode. As discussed in the main text, generalization to multiple emitters is straightforward. For illustration, we begin with a Jaynes-Cummings model 
\be
H_{\rm JC}= \nu_{\rm bare} \sigma^z/2 + \omega_{\rm bare}  a^\dag a + \frac{g}{2} (\sigma^- a^\dag + \sigma^+ a), 
\label{eq:JaynesCummings}
\ee
where $\sigma^{x,y,z}$ are Pauli matrices, $\sigma^\pm = (\sigma^x \pm i \sigma^y)/2$, and $g = \mathcal{M} E_0$ is the coupling constant that captures the dipole interaction between the exciton mode (described by its dipole moment $\mathcal{M}$) and the cavity mode (described by the amplitude of the electric field $E_0$). Here $\nu_{\rm bare}$ and $\omega_{\rm bare}$ are the bare frequencies of the exciton and cavity mode resonances in the absence of any interaction, i.e., $g=0$. The interaction -- i.e., the third term of Eq.~(\ref{eq:JaynesCummings}) -- is responsible for the optical Stark effect. To see this 
we perform a canonical transform $T^\dag H_{\rm JC} T$ with $T = \exp{S}$; here $S= -g (a  \sigma^+ - a^\dagger \sigma^-)/ (2\Delta)$. The transformed Hamiltonian is given by 
\be \label{tildeH}
T^\dag H_{\rm JC} T = H_{\rm JC} + [H_{\rm JC},S] + \frac{1}{2!} [[H_{\rm JC},S],S] + \frac{1}{3!} [[[H_{\rm JC},S],S],S] + \cdots . 
\ee
We note that $[\nu_{\rm bare} \sigma^z/2 + \omega_{\rm bare}  a^\dag a, S] = - g(\sigma^- a^\dag + \sigma^+ a) /2$. 
By taking the dispersive limit $g \ll \Delta$, the transformed Hamiltonian up to the first order in $g / \Delta$ is 
\begin{align}
T^\dag H_{\rm JC} T & = \omega_{\rm bare} a^\dagger a + \nu_{\rm bare}  \sigma^z/2 + V a^\dagger a \vec \sigma^z/2 + V \sigma^z/4 + V/4 + \mathcal{O}(V^2),  \nonumber \\
& = (\omega_{\rm bare} - V/2) a^\dag a  + (\nu_{\rm bare} + V/2)  \hat{P} + V a^\dag a \hat{P} - \nu_{\rm bare}/2 + \mathcal{O}(V^2), 
\label{eq:transformed}
\end{align}
where $V =  g^2/(2\Delta)$ is the dispersive coupling constant and $\hat{P} = \sigma^z/2 + \mathbbm{1}/2$. We note that the last term, $- \nu_{\rm bare}/2 $, is a constant offset to the Hamiltonian that does not play any role in the physics we discuss. As a result, in the main text as well as what follows, we drop mention of the constant offset. Using this transformation, Eq.~(\ref{H one-valley}) can be readily read off Eq.~(\ref{eq:transformed}).

\subsection{Stability analysis of the steady state solutions}
In the plots in the main text, we have shown the multiple branches of steady state solutions that appear at a given set of drive parameters. In these plots, solid lines denote the stable solutions while dashed lines denote unstable solutions. The stability of each steady state solution can be ascertained by examining the time-evolution of the exciton occupation factor $\partial_t \la \hat{P} (t)\ra =  \partial_t \la s^z (t)\ra$ in Eq.~(\ref{eq:sdynamics}). In particular, taking a small deviation $\delta P$ from the steady state solution $\overline{P}$ and computing $\partial_t \la \hat{P} (t)\ra \vert_{ \overline{P} + \delta P}$, we find a solution is stable if $\rm{sgn} \left( \partial_t \la \hat{P} (t) \ra \vert_{\overline{P} + \delta P} \right) = -\rm{sgn} (\delta P)$; i.e., the solution is stable if, after the system is slightly displaced from the steady state, it time evolves back to the original steady state solution $\la \hat{P} (t) \ra = \overline{P}$.

\subsection{Steady state exciton population with multiple emitters} 

In this section, we describe the steady state solution for the MSE with multiple emitters. As in the main text, we replace 
$\hat{P} \to \hat{P}_{\rm tot} = \sum_{i} \hat{P}_i = \sum_{i} (s^z_i + 1/2)$ in the Hamiltonian, Eq.~(\ref{H one-valley}), as well as $\sigma^{+,-} \to s^{+,-}_{\rm tot} = \sum_{i} \sigma_{i}^{+,-}$ in Eq.~(\ref{eq:driven H}), where the sum runs over the degenerate excitonic emitters. Similarly, $\tilde{\omega} (P) \to \tilde{\omega} = \omega^{(0)} + V P_{\rm tot}$ in Eq.~(\ref{eq:mutualtuning}) where $P_{\rm tot} = 0 , 1, 2,  \cdots $ are eigenvalues of $\hat{P}_{\rm tot}$. Since all the emitters interact with the same cavity photon mode, $\tilde{\nu}(m)$ in Eq.~(\ref{eq:mutualtuning}) remains unchanged. For simplicity of notation, we have defined the spin operators $\vec s = \boldsymbol{\sigma}/2$, where $\boldsymbol{\sigma} = (\sigma^x, \sigma^y, \sigma^z$) is the vector of Pauli matrices.

Similar to our approach described for a single excitonic emitter in the main text, we exploit a separation of time scales between cavity and exciton systems that is  easily achieved in readily available systems (see discussion in main text). In this fashion, we can mean-field decouple the excitonic and cavity degrees of freedom. As a result the 
excitonic system  evolves according to
\be
\partial_t \tilde{\rho}_{X, {\rm tot}} (t) = i [ \tilde{\rho}_{X, {\rm tot}}  (t), - \delta \nu (t) \hat{P}_{\rm tot} + F_X s_{\rm tot}^{x}] + \mathcal{I}_X [\tilde{\rho}_{X, {\rm tot}}  (t)],
\label{eq:mastermultiple}
\ee
where 
$\delta \nu(t) = \nu_d - \tilde{\nu}[\la m(t) \ra]$, and the 
dissipator for the multiple emitters reads
\be
\mathcal{I}_X [\tilde{\rho}_{X, {\rm tot}} ] =  \gamma \big[2 s^{-}_{\rm tot}   \tilde{\rho}_{X, {\rm tot}}   s^{+}_{\rm tot} - s^{+}_{\rm tot}s^{-}_{\rm tot}\tilde{\rho}_{X, {\rm tot}} - \tilde{\rho}_{X, {\rm tot}} s^{+}_{\rm tot}s^{-}_{\rm tot} \big].
\ee
As a sanity check, noting that $ [\tilde{\rho}_{X, {\rm tot}} , \mathbbm{1} ] = 0$, we find that probability is conserved: $\partial_t {\rm Tr} ( \tilde{\rho}_{X, {\rm tot}}  ) = 0$. 

We can use Eq.~(\ref{eq:mastermultiple}) to obtain the equation of motions governing the ``Bloch'' (giant spin) dynamics of the multiple emitters. Here we treat the collection of multiple emitters as a giant spin with magnitude $\mathcal{S} = \mathcal{N}/2$. When the spin is pointing down such that $\la s^z_{\rm tot} \ra = -\mathcal{S}$, no excitons are excited. When $\la s^z_{\rm tot} \ra = \mathcal{S}$, all the excitonic modes are excited. In this work we focus on the regime far from saturation, where the exciton density remains low.

Writing 
$\la s^{i}_{\rm tot} (t) \ra = {\rm Tr} [s^i_{\rm tot}\tilde{\rho}_{X, {\rm tot}}(t)]$ for $i=x,y,z$, we find
\begin{align} 
\partial_t \la s^{x}_{\rm tot} (t)\ra & =  \delta \nu (t) \la s^{y}_{\rm tot} (t) \ra - \gamma \la s^{x}_{\rm tot} (t) \ra + \gamma \la \{ s^{z}_{\rm tot} ,s^{x}_{\rm tot} \}\ra, \nonumber \\ 
\partial_t \la s^{y}_{\rm tot} (t)\ra & = -\delta \nu (t) \la s^{x}_{\rm tot} (t) \ra - F_X \la s^{z}_{\rm tot} (t) \ra - \gamma \la s^{y}_{\rm tot} (t) \ra 
 + \gamma \la \{ s^{z}_{\rm tot} ,s^{y}_{\rm tot} \}\ra, \nonumber \\
\partial_t \la s^{z}_{\rm tot} (t)\ra & = F_X \la s^{y}_{\rm tot} (t) \ra - 2\gamma \big[ \la s_{\rm tot}^2 \ra - \la [s^z_{\rm tot}]^2 \ra + \la s^{z}_{\rm tot} (t)\ra \big],
\label{eq:GiantSpinEOM}
\end{align}
where $\la s_{\rm tot}^2 \ra = \la  [s^{x}_{\rm tot}]^2 + [s^{y}_{\rm tot}]^2  + [s^{z}_{\rm tot}]^2 \ra = \mathcal{S}(\mathcal{S}+1)$  
and $\{ a, b\} = ab + ba $ is the  anti-commutator. In obtaining Eq.~(\ref{eq:GiantSpinEOM}) we have cycled the operators in the trace, recalling that ${\rm Tr} (\widehat{\mathcal{O}}_A \widehat{\mathcal{O}}_B \widehat{\mathcal{O}}_C ) = {\rm Tr} (\widehat{\mathcal{O}}_C \widehat{\mathcal{O}}_A \widehat{\mathcal{O}}_B )  = {\rm Tr} (\widehat{\mathcal{O}}_B \widehat{\mathcal{O}}_C \widehat{\mathcal{O}}_A )$, as well as noted the identity $[s^{i}_{\rm tot}, s^{j}_{\rm tot}] = i \epsilon_{ijk} s^{k}_{\rm tot}$, where $\epsilon_{ijk}$ is the Levi-Civita symbol. 

\subsubsection{Spin coherent state ansatz} 

As we now discuss, the dynamics of the multiple emitters can be understood as the dynamics of a giant spin $\hat{\vec s}_{\rm tot} = s_{\rm tot}^x \hat{\vec x} + s_{\rm tot}^y\hat{\vec y}+ s_{\rm tot}^z \hat{\vec z}$ with a spin magnitude $\mathcal{S}$. Here $\mathcal{S} = \mathcal{N}/2$ can be estimated from the degeneracy of the multiple emitters in the system. Take for example the limiting case discussed in detail in the main text: a single two level system (i.e., spin $\mathcal{S} = 1/2$ system). For spin-1/2, we recall that $\{s^z, s^y \} = \{s^x, s^y \} = \{ s^z, s^x\} =0$; similarly, $\la s^2 \ra  - \la (s^z)^2\ra  = 3/4 - 1/4 = 1/2$. Using these identities, we find Eq.~(\ref{eq:GiantSpinEOM}) reduces to Eq.~(\ref{eq:sdynamics}) of the main text. However when multiple emitters are involved, $\{ s_{\rm tot}^z ,s_{\rm tot}^x \}$ does not necessarily vanish. 

Key to analyzing Eq.~(\ref{eq:GiantSpinEOM}) is obtaining (closed) forms for the terms $ \la \{ s_{\rm tot}^z ,s_{\rm tot}^y\}\ra, \la \{ s_{\rm tot}^z ,s_{\rm tot}^x \}\ra$, and  $\la s_z^2 \ra $ in terms of the giant spin expectations values $\la s^{i}_{\rm tot}\ra$. In order to do so and find the steady-state solutions of Eq.~(\ref{eq:GiantSpinEOM}), we adopt a spin coherent state ansatz 
\be
\tilde{\rho}_{X, {\rm tot}} = | \psi_{\hat{\vec n}} \ra \la \psi_{\hat{\vec n}} |, \quad \vec s_{\rm tot}  \cdot \hat{\vec n} | \psi_{\hat{\vec n}} \ra = \mathcal{S} | \psi_{\hat{\vec n}} \ra, 
\ee
with the giant spin pointing in the direction $\hat{\vec n}$: 
\be
\hat{\vec n}  = \frac{1}{\mathcal{S}} \Big(\la s_{\rm tot}^x \ra, \la s_{\rm tot}^y\ra, \la s_{\rm tot}^z\ra \Big). 
\ee
In using this coherent state, it is useful to note that expectation values of various operators (e.g., $ s_{\rm tot}^z  s_{\rm tot}^x$) in this coherent state can be obtained by applying a suitable rotation to the operator and evaluating the expectation value in a well-known reference state (e.g., $|\psi_{\hat{\vec z}} \ra$). For example, 
\be
\mathcal{S} =  \la  \psi_{\hat{\vec n}}| \vec s_{\rm tot}  \cdot \hat{\vec n} | \psi_{\hat{\vec n}} \ra = \la  \psi_{\hat{\vec z}}| U\big(\vec s_{\rm tot}  \cdot \hat{\vec n}\big)U^\dag | \psi_{\hat{\vec z}} \ra, \quad U\big(\vec s_{\rm tot}  \cdot \hat{\vec n}\big)U^\dag = s_{\rm tot}^z, \quad U | \psi_{\hat{\vec n}} \ra =  | \psi_{\hat{\vec z}} \ra, 
\ee
where $U$ is unitary operator that rotates the spin (determined by $\hat{\vec n}$). 

Using this coherent state ansatz, we obtain
\begin{align}
\la [s^z_{\rm tot}]^2 \ra &= s_\perp^2/\big(2\mathcal{S}\big)+ \la s^z_{\rm tot}\ra ^2, \nonumber \\
\la  \{ s^z_{\rm tot}, s^x_{\rm tot} \}  \ra &= - \la s^z_{\rm tot}\ra \la s^x_{\rm tot}\ra/\mathcal{S} + 2\la s^z_{\rm tot}\ra \la s^x_{\rm tot} \ra, \nonumber \\
\la  \{s^z_{\rm tot},  s^y_{\rm tot} \}  \ra &= - \la s^z_{\rm tot}\ra \la s^y_{\rm tot}\ra/\mathcal{S} + 2\la s^z_{\rm tot}\ra \la s^y_{\rm tot}\ra, 
\label{eq:szsx}
\end{align}
where $s_\perp = ( \la s^x_{\rm tot} \ra^2 + \la s^y_{\rm tot} \ra^2 )^{1/2}$. In obtaining Eq.~(\ref{eq:szsx}) we have used the identity $\la  \psi_{\hat{\vec z}}| s^x_{\rm tot} s^z_{\rm tot} | \psi_{\hat{\vec z}}\ra = \la  \psi_{\hat{\vec z}}| s^y_{\rm tot} s^z_{\rm tot} | \psi_{\hat{\vec z}}\ra  =0$. Further we have noted that  $\la  \psi_{\hat{\vec z}}| s^x_{\rm tot} s^y_{\rm tot} |  \psi_{\hat{\vec z}}\ra = - \la  \psi_{\hat{\vec z}}| s^y_{\rm tot} s^x_{\rm tot} |  \psi_{\hat{\vec z}}\ra $. This latter identity can be discerned from noting that in $s^x_{\rm tot} s^y_{\rm tot} = \sum_{ij} s^x_i s^y_j$ only the terms where $i=j$ yield non-zero values, and that when $i=j$, we have $\{ s^x,s^y\} =0$. 
Substituting Eq.~(\ref{eq:szsx}) into Eq.~(\ref{eq:GiantSpinEOM}) we get
\begin{align}
\partial_t \la s_{\rm tot}^x (t)\ra & =  \delta \nu (t) \la s_{\rm tot}^y(t) \ra - \gamma \la s_{\rm tot}^x (t) \ra \big[1+ \la s^z_{\rm tot}\ra/\mathcal{S} - 2\la s^z_{\rm tot}\ra\big], \nonumber \\ 
\partial_t \la s_{\rm tot}^y(t)\ra & = -\delta \nu (t) \la s_{\rm tot}^x (t) \ra - F_X \la s_{\rm tot}^z (t) \ra - \gamma \la s_{\rm tot}^y(t) \ra  \big[1+ \la s^z_{\rm tot}\ra/\mathcal{S} - 2\la s^z_{\rm tot}\ra\big] \nonumber \\
\partial_t \la s_{\rm tot}^z (t)\ra & = F_X \la s_{\rm tot}^y(t) \ra - 2\gamma \big[ \mathcal{S}(\mathcal{S}+1) -  s_\perp^2/(2\mathcal{S}) - \la s^z_{\rm tot}\ra ^2 + \la s_{\rm tot}^z (t)\ra \big].
\label{eq:GiantSpinEOMCoherent}
\end{align}

In the absence of a drive, $F_X = 0$, and at steady state, the excitonic system yields its equilibrium value with $\la s^z_{\rm tot}\ra^{\rm eq} = - \mathcal{S}$ (similarly, $\la s^x_{\rm tot}\ra^{\rm eq} = \la s^y_{\rm tot}\ra^{\rm eq} = 0$). Interestingly, Eq.~(\ref{eq:GiantSpinEOMCoherent}) is a non-linear equation in $\la s^z_{\rm tot}\ra$. Such nonlinearities become important only when the deviation of $\la s^z_{\rm tot}\ra$ away from its equilibrium (no excitation) value is large (i.e., comparable to $\la s^z_{\rm tot}\ra^{\rm eq}$), due to saturation of the exciton resonance. 

We now analyze the steady state population of the excitons by first taking $\tilde{\nu} [\la m(t) \ra] \to \tilde{\nu} (m)$ for a fixed $m$ in the same way as discussed in the main text. In the low-excitation regime 
where $\la s^z_{\rm tot}\ra = -\mathcal{S} + \overline{P}_{\rm tot}$ for $\overline{P}_{\rm tot} \ll \mathcal{S}$, we can safely linearize Eq.~(\ref{eq:GiantSpinEOMCoherent}) in $\overline{P}_{\rm tot}$. Solving for $\overline{P}_{\rm tot}$ at steady state in Eq.~(\ref{eq:GiantSpinEOMCoherent}), we obtain  
\be
\overline{P}_{\rm tot} = \frac{\mathcal{S} F_X^2 }{F_X^2+ 2(\Gamma^2 + [\nu_d- \nu(m)]^2)}, 
\label{eq:ptot}
\ee
where $\Gamma \sim 2\mathcal{S} \gamma$ is the recombination rate for excitons in the entire multiple emitter system. In obtaining Eq.~(\ref{eq:ptot}), we have used that $s_\perp^2/\mathcal{S} \ll \overline{P}_{\rm tot}$. 
Writing $\mathcal{S} = \mathcal{N}/2$ produces Eq.~(\ref{eq:stot}) in the main text. 

\subsubsection{Alternative derivation of exciton population in the low-density limit: Holstein-Primakoff transformation}

In this section, we provide an alternative derivation of the steady state exciton population (in the case of multiple degenerate emitters) in the low-density limit. In so doing, we note that the dynamics of the exciton population can be understood from analyzing the evolution of a large spin with magnitude $\mathcal{S}\gg 1$. In the large $\mathcal{S}$ limit, the excitations of this system can be analyzed in a bosonic framework, using a Holstein-Primakoff transformation. To begin, we write 
\be
s^+_{\rm tot} \to (\sqrt{2\mathcal{S}} ) b^\dag \Big[1-\frac{b^\dag b}{2\mathcal{S}}\Big]^{1/2}, 
\quad s^-_{\rm tot} \to (\sqrt{2\mathcal{S}} ) \Big[1-\frac{b^\dag b}{2\mathcal{S}}\Big]^{1/2} b, 
\label{eq:HP1}
\ee
where $b^\dag$ is a bosonic creation operator so that $[b,b^\dag] = 1$. 
Using Eq.~(\ref{eq:HP1}) we can readily verify
\be
\label{eq:Sz}2s^z_{\rm tot} = [s^+_{\rm tot}, s^-_{\rm tot}] = 2\mathcal{S}\Big[b^\dag \big(1-b^\dag b/(2\mathcal{S})\big)^{1/2},\big(1-b^\dag b/(2\mathcal{S})\big)^{1/2} b\Big]= 2 (b^\dag b - \mathcal{S}).
\ee
Note that Eq.~(\ref{eq:Sz}) is exact; the factors of $[1-\frac{b^\dag b}{2\mathcal{S}}]^{1/2}$ must be included in order to obtain the $b^\dagger b$ dependence on the right-hand side (which will prove to be essential below).

Noting that $b^\dagger b$ counts the number of excitons created [see Eq.~(\ref{eq:Sz})] we 
express the state where $P_{\rm tot}$ excitons have been excited as 
\be
\label{eq:state} |P_{\rm tot}\ra \propto \big[b^\dag\big]^{P_{\rm tot}}|0\ra_{\rm boson}, 
\quad P_{\rm tot} < 2\mathcal{S}, 
\ee
where the vacuum state (no excitons present) corresponds to the giant spin pointing downwards ($\la s^z_{\rm tot}\ra = -\mathcal{S}$).

In the co-rotating frame of the drive, we find the Hamiltonian that describes the excitons is 
\be
\tilde{H}_{X, {\rm tot}} = (\tilde{\nu}
- \nu_d) \hat{P}_{\rm tot} + F_X s^x_{\rm tot}  \overset{{\rm HP}}{\longrightarrow} (\tilde{\nu}
- \nu_d) b^\dag b + F_X (\sqrt{2\mathcal{S}})( b + b^\dag) /2,
\label{eq:HP2} 
\ee
where we have recalled that $\hat{P}_{\rm tot} = s^z_{\rm tot} + \sum_i 1/2 \approx s^z_{\rm tot} + \mathcal{S}$ and $s_{\rm tot}^x = (s^+_{\rm tot} + s^-_{\rm tot} )/2$.
Here we have estimated $\mathcal{N} = 2\mathcal{S}$; the arrow labeled HP denotes the Holstein-Primakoff transformation. In obtaining the last term of Eq.~(\ref{eq:HP2}) we have taken the limit $ \la b^\dag b\ra \ll \mathcal{S}$ so that $s^+_{\rm tot} \approx (\sqrt{2\mathcal{S}}) b^\dag$ and $s^-_{\rm tot} \approx (\sqrt{2\mathcal{S}}) b$.  

Similarly, we find the exciton-cavity interaction reads as
\be
\tilde{H}_{\rm int, \rm tot} = V a^\dag a \hat{P}_{\rm tot} \overset{{\rm HP}}{\longrightarrow} V a^\dag a b^\dag b. 
\ee
This expression captures the nonlinear interaction between the exciton degrees of freedom (characterized by $b$'s) and the cavity photons (characterized by $a$'s).
Note that this interaction is distinct from the bilinear exciton-cavity interaction typically 
discussed for the polariton effect. 
The nonlinearity arises due to the underlying fermionic nature of the electrons and holes that make up the excitons~\cite{Combescot}. 
  The role of the fermionic Pauli exclusion is reflected in the square root factors in the relation between the collective exciton creation operator $s_{\rm tot}^+$ and the bosonic creation operator $b^\dagger$ (and similarly for $s_{\rm tot}^-$ and $b$), as well as the $b^\dagger b$ dependence of $[s^+_{\rm tot}, s^-_{\rm tot}]$ as shown in Eq.~(\ref{eq:Sz}).

In a similar fashion to that described in the main text and above, we exploit a separation of time scales between the exciton relaxation and cavity relaxation time scales and write the equation of motion of the excitonic excitations as 
\be
\partial_t \tilde{\rho}_{X,{\rm tot}}^{\rm HP} (t) = i [ \tilde{\rho}_{X,{\rm tot}}^{\rm HP}  (t), (\tilde{\nu}
- \nu_{\rm d} ) b^\dag b + F_X \sqrt{\mathcal{S}/2}( b + b^\dag) ] + \mathcal{I}_X^{\rm HP} [\tilde{\rho}_{X,{\rm tot}}^{\rm HP}  (t)],
\label{eq:masterHP}
\ee
where $
\mathcal{I}_X^{\rm HP} [\tilde{\rho}] = \Gamma (2 b\tilde{\rho} b^\dagger - b^\dagger b\tilde{\rho}  -  \tilde{\rho} b^\dagger b )$, with $\Gamma$ capturing a phenomenological decay rate of the excitonic mode. 

The steady state solution to Eq.~(\ref{eq:masterHP}) can be readily obtained using a coherent state ansatz $\tilde{\rho}_X^{\rm tot, HP} = | \beta \ra \la \beta |$ where $b |\beta \ra = \beta |\beta \ra$ and $\beta \in \mathbb{C}$. 
By direct calculation we verify that the coherent state ansatz indeed yields a steady state solution of Eq.~(\ref{eq:decoupled2}) for fixed oscillator excitation $m$, with $\beta = (- iF_X \sqrt{\mathcal{S}/2})/(\Gamma - i (\nu_{\rm d} - i\tilde \nu [m]))$.
The corresponding steady state exciton population is given by
\be
\label{eq:PtotHP}
\overline{P}_{\rm tot} \approx \la b^\dag b\ra = (\mathcal{S}F_X^2/2)/\{\Gamma^2 + \big(\nu_{\rm d} - \tilde \nu [m ]\big)^2\}.
\ee
As a consistency check, we note that in the limit $F_X \ll \Gamma, \nu_d, \tilde{\nu}$, Eq.~(\ref{eq:ptot}) reduces to Eq.~(\ref{eq:PtotHP}).

\subsection{Estimate of parameter values for MSE in a TMD photonic cavity} 

In this section, we estimate the values of the parameters used in the main text to achieve the MSE in a transition metal dichacolgenide (TMD) sample placed in a photonic crystal cavity. For the purposes of the estimates below, we will restore $\hbar$ so that $\hbar \omega, \hbar \nu$ as well as $\hbar \kappa, \hbar \Gamma$ are energies. 

We consider a cavity with a high quality factor $Q = \omega^{(0)} / \kappa $ and compressed 
effective mode volume $V_{\rm mode} = \alpha (\lambda/n)^3$, where $\kappa$ is the cavity mode linewidth, $\omega^{(0)}$ is the bare cavity resonance frequency, $\lambda$ is the free space wavelength of the photons, 
$n$ is the refractive index of the cavity, and $\alpha$ is a proportionality constant that depends on the geometry of the cavity. In photonic crystal cavities, $\alpha$ values in the range $0.02 \sim 1$ have been achieved with quality factors of order $10^3 \sim 10^6$~\cite{PhC1,PhC2, PhC3, PhC4}. Here for the purposes of a simple demonstration, we choose $\alpha = 0.05$ and $Q= 20000$. Similarly, we will consider an exciton resonance at $\hbar \nu^{(0)} = 2 \, {\rm eV}$ and set the cavity mode resonance frequency, $\omega^{(0)}$, to be red-detuned away from it (as discussed below, we choose a detuning of $43 \, {\rm meV}$). These parameters correspond to cavity linewidth of $ \hbar \kappa \approx 0.1 \, {\rm meV}$; taking the refractive index in the cavity as $n \approx 3$, we obtain an effective mode volume of $V_{\rm mode} \approx 4.7 \times 10^5$ nm$^3$.

\subsubsection{TMD Stark shift and exciton-cavity interaction}

Here we estimate the value for the dispersive coupling $V$ for a TMD material placed on the photonic crystal cavity described in the main text. First, we note that excitons in monolayer TMDs obey circularly polarized optical selection rules wherein the excitons around $K$ ($K'$) valleys primarily interact with photons of left (right) circular polarization. This valley-circularly polarized light selectivity justifies our use of single flavor of exciton modes (e.g., excitons in the $K$ valley) interacting with a single cavity mode (e.g., left hand circularly polarized cavity mode) in the main text. 
Indeed, this selectivity is well evidenced in experiment. It not only manifests in selective excitation of excitons in the valleys (by using circularly polarized light), but also yields a valley Stark effect wherein light of left (right) circular polarization only shifts the exciton resonances of $K$ ($K'$) excitons~\cite{stark, stark_BS, stark_biexciton}. For a left-hand circularly polarized field $\vec E (t) = E_0 (\hat{\vec x} \cos \omega t + \hat{\vec y} \sin \omega t)$, excitons in the $K$ valley experience a Stark shift~\cite{stark, stark_BS, stark_biexciton}
\be
{\rm Stark} \,\, {\rm shift} = \delta E = (\mathcal{M} E_0)^2/2\Delta \propto E_0^2/\Delta,
\label{eq:exptstark}
\ee
where $\Delta =\hbar( \nu^{(0)} - \omega )$ and $\mathcal{M}$ is a material dependent dipole matrix element for the circularly polarized field. 

In Ref.~\cite{stark}, an optical Stark shift in a single valley of monolayer WS$_2$ was measured. Ref.~\cite{stark} reported that a peak Stark shift value $\delta E_{\rm expt} = 18\,{\rm meV}$ was induced by a circularly polarized laser pulse with fluence of 120 $\mu$J/cm$^2$ and pulse width 160 fs. The detuning between the exciton and laser pulse was $\Delta_{\rm expt} = 180$ meV. Recalling that the intensity of a circularly polarized field $\vec E(t)$ (see above) is $I = \epsilon_0 c |E_0^{\rm expt}|^2$, we estimate peak $E_0^{\rm expt} = 5.3 \times 10^7$ V/m in Ref.~\cite{stark}, where $\epsilon_0$ is the vacuum permittivity, and $c$ is the speed of light. Here we have estimated peak intensity as fluence$/$pulse width. 

The value of $V$ (Stark shift per circularly polarized photon in the cavity) for our TMD in a photonic cavity setup can be estimated by comparing with the above experiments on the Stark effect in WS$_2$~\cite{stark}. To do so, we first note that the electric field amplitude of the circularly polarized cavity mode can be estimated as $\hbar \omega^{(0)} = V_{\rm mode} \epsilon_0 \epsilon_{\rm cav}^{\rm eff} [E_{\rm cav}^{(0)}]^2$, where $\epsilon_{c}^{\rm eff}$ is the effective relative permittivity of the cavity. Taking $V_{\rm mode}$ from above, and $\epsilon_{\rm cav}^{\rm eff} =10$, we estimate an electric field strength of the circularly polarized cavity mode as ${E}_{\rm cav}^{(0)}=2.7 \times 10^6$ V/m. Comparing the cavity mode electric field amplitude with that used in the experiment above~\cite{stark} and specifying an exciton-cavity detuning $\Delta_{\rm cav} = 43\, {\rm meV}$, we obtain $V$ (Stark shift per circularly polarized photon in the cavity) as
\be
V = \frac{(E_{\rm cav}^{(0)}/E_{\rm expt}\big)^2}{\Delta_{\rm cav}/\Delta_{\rm expt}} \times \delta E_{\rm expt} \approx 0.2 \, {\rm meV}, 
\label{eq:Vestimate}
\ee
where we have used the proportionality relation in latter part of Eq.~(\ref{eq:exptstark}).
While we have chosen a specific value of $\Delta_{\rm cav}$ which yields $V \approx 0.2 \, {\rm meV}$, a range of values of $V \sim 0.1 - 0.5 \, {\rm meV}$ can be readily achieved by tuning $\Delta_{\rm cav}$.  

We note that $V$ in the cavity system can also be estimated directly from the dipole matrix element for the circularly polarized 
field acting on a TMD (in our case, e.g., WS$_2$). Using Eq.~(\ref{eq:exptstark}) and the experimental parameters from Ref.~\cite{stark} discussed above, we find $\mathcal{M} \approx 73.6 \, {\rm Debye}$. We note that this experimentally inferred $\mathcal{M}$ is within a factor of unity from that obtained from a simple theoretical estimate~\cite{stark} $\mathcal{M}_{\rm theoretical} \approx 56\, {\rm Debye}$. For the exciton-cavity photon system, we recall that the exciton-cavity photon interaction in Eq.~(\ref{eq:JaynesCummings}) is $g = \mathcal{M}E_{\rm cav}^{(0)} \approx 4.15\, {\rm meV}$. Using $V=g^2/2\Delta_{\rm cav}$ from Eq.~(\ref{eq:transformed}) we obtain $V \approx 0.2 \, {\rm meV}$ in agreement with Eq.~(\ref{eq:Vestimate}).

\subsubsection{Exciton drive strength, $F_X$}

In this section, we connect the exciton driving field intensity (in kW/cm$^2$) to the $F_X$ driving strength (in meV) used in the main text.
The strength of the exciton drive $F_X$ in Eq.~(\ref{eq:driven H}) of the main text can be obtained in the same fashion as that described above via $F_X =  \mathcal{M} E^{\rm d}_X$, where $E^{\rm d}_X$ is the amplitude of a circularly polarized driving electric field $E_{X}^{\rm drive} (t) = E_X^{\rm d} (\cos \nu_d t \,\hat{\vec{x}} + \sin \nu_d t \,\hat{\vec{y}})$. 
The intensity of the (exciton) driving field is given by 
$I_X = \epsilon_0 \epsilon^{\rm eff} c (E^{\rm d}_X)^2 $, where $\epsilon^{\rm eff}$ is an effective relative permittivity that depends on the geometry of the incident irradiation. For illustration, we have taken $\epsilon^{\rm eff}=1$. For a driving intensity of $I_X = {0.1} - 0.6$ kW/cm$^2$, we estimate a circularly polarized driving electric field strength as $E^{\rm d}_X = {0.1 -} 4.8 \times 10^4$ V/m. Using the value of $\mathcal{M}$ above we obtain 
$F_X \approx {0.002 -} 0.072$ meV. We note parenthetically that these values of drive electric field are far smaller (two orders of magnitude smaller) than the electric field amplitude of the cavity mode discussed above. 

\subsubsection{Exciton recombination rate}

The recombination rate of excitons in TMDs is typically in the range of $0.2 \sim {\rm several}$ ps at cryogenic temperatures \cite{WS2_lifetime, WSe2_lifetime, exciton_lifetime_2ps, exciton_lifetime, MoS2_hBN}. To illustrate MSE, we used $\hbar \Gamma = 1$ meV.

\subsubsection{Cavity drive strength, $F_0$} 

In this subsection, we relate the incident cavity drive power $P$ to the cavity drive parameter $F_0$ used in the main text. 
To do so we first briefly review classical coupled mode theory~\cite{Joannopoulos}. 
  In a cavity with cavity eigenmodes $\vec e_{i}(\vec r)$, the electric field profile in the cavity can be described by $\vec E^{\rm cavity} (\vec r,t)=  {\rm Re} \big[\sum_iC_i (t) \vec e_{i}(\vec r)\big]$; here $i$ indexes the cavity eigenmodes. $C_i (t)$ is a time-varying amplitude that describes the degree to which each of the cavity eigenmodes are occupied over time. 
According to coupled mode theory, the classical cavity mode fields coupled to an external drive evolve according to~\cite{Joannopoulos} 
\be
{\frac{dC_i(t)}{dt} = - i \omega_i C_i(t) - \kappa C_i(t) - i \xi f_i^+(t), }
\label{eq:classicalCMT}
\ee
where $f_i^+(t) $ is the amplitude of the incident drive field that couples to the cavity mode $i$.
Here $\omega_i$ is the resonance frequency of cavity mode $i$, $\kappa$ is the decay rate of the cavity mode, and $\xi = \sqrt{2 \kappa}$ represents the strength of the cavity-incident driving field coupling~\cite{Joannopoulos}. 

We now focus on a planar photonic cavity where there are two degenerate modes in the plane, which we label $\vec e_x$ and $\vec e_y$, with common frequency $\omega_x = \omega_y = \omega$.
  Here $\vec e_x$ and $\vec e_y$ are cavity modes that are polarized in the $x$ and $y$ directions.
  We will also assume that the system 
  is isotropic in-plane so that $\kappa_x = \kappa_y = \kappa$.
  Further, we focus on cavities that primarily decay radiatively, 
  consistent with a high quality factor cavity. 
We remark that, following the standard convention~\cite{Joannopoulos}, the cavity modes profiles $\vec e_{x,y} (\vec r)$ can be normalized in such a way
that 
\be
U(t) = \sum_i |C_i(t)|^2 = |C_x(t)|^2 + |C_y (t)|^2,  \quad \mathcal{P} = \sum_i |f_i^+ (t) |^2 = |f_x^+ (t) |^2 + |f_y^+ (t) |^2,
\label{eq:cavityenergy}
\ee 
where $U(t)$ is the energy in the cavity and $\mathcal{P}$ is the incident driving power. 
To describe the driving of the cavity by a circularly polarized beam we will use $f_x^+ (t)  = f_0 \cos \omega_d t$ and $f_y^+ (t)  = f_0 \sin \omega_d t$. 

We now turn to the quantum mechanical description of the driven cavity:
\be
H_{\rm cav} = \hbar \omega (a_x^\dag a_x + a^\dag_y a_y) + \frac{F_0}{\sqrt{2}} (a_x^\dag + a_x) \cos \omega_dt +  \frac{F_0}{\sqrt{2}}(a_y^\dag + a_y) \sin \omega_dt, 
\label{eq:cavityhamiltonian}
\ee
where $a_{x,y}$ are annihilation operators for the $x,y$ polarization cavity photon modes (in modes $\vec{e}_{x,y}$), and the last two terms describe a circularly polarized driving with amplitude $F_0$. The latter terms describe the full Rabi-type drive-cavity coupling, see below for discussion.  Here we have dropped the vacuum energy of the cavity mode as it acts as a constant energy off-set; including it does not alter our conclusions/estimate for $F_0$ below.

In the same fashion as described in the main text, the evolution of the cavity can be described by the master equation
\be
\partial_t \rho_{\rm cav} (t)  = \frac{i}{\hbar} [\rho_{\rm cav} (t), H_{\rm cav}] - \mathcal{I} [\rho_{\rm cav} (t)], 
\label{eq:masterequationcavity}
\ee
where $\mathcal{I} (\rho)= \kappa\big[a_x^\dag a_x \rho - 2 a_x \rho a_x^\dag + \rho a_x^\dag a_x + (x\to y)\big]$ tracks the decay of the mode.

In order to track the evolution of the amplitudes above, 
we 
define the mode amplitude operators $\widehat{C}_i = C_i^{(0)} a_i $, for $i = x,y$). 
Following Eq.~(\ref{eq:cavityenergy}) above, we have $U  = \sum_i \la \widehat{C}_i^\dag \widehat{C}_i \ra = |C_x^{(0)}|^2  \la a_x^\dag a_x \ra +|C_y^{(0)}|^2  \la a_y^\dag a_y \ra $. 
This energy must correspond to the energy of the cavity obtained directly from the first term of Eq.~(\ref{eq:cavityhamiltonian}): $\hbar \omega \la a_x^\dag a_x + a^\dag_ya_y \ra $. As a result, we obtain $C^{(0)} = |C_x^{(0)}| = |C_y^{(0)}| = \sqrt{\hbar \omega}$. 

The temporal evolution of the cavity amplitude $ \la \widehat{C}_i(t)\ra  = {\rm Tr} \big[ \widehat{C}_i \rho_{\rm cav} (t) \big]$ can be described by multiplying $\widehat{C}_i$ into Eq.~(\ref{eq:masterequationcavity}) and taking the trace. Recalling the commutator identities $[a^\dag a, a] = - a$, $[a,a^\dag] = 1$, and cyclically permuting the operators, we obtain 
\be
\frac{d\la \widehat{C}_x(t)\ra}{dt} = -i\omega \la \widehat{C}_x(t)\ra - \kappa \la\widehat{C}_x(t)\ra  - i \frac{F_0C^{(0)}}{\sqrt{2} \hbar} {\rm cos} \omega_dt, \quad \frac{d\la \widehat{C}_y(t)\ra }{dt} = -i\omega \la \widehat{C}_y(t)\ra - \kappa \la\widehat{C}_y(t)\ra  - i \frac{F_0C^{(0)}}{\sqrt{2} \hbar} {\rm sin} \omega_dt.
\label{eq:quantumCMT}
\ee
Relating $\avg{\widehat{C}_i(t)}$ with $C_i(t)$ in Eq.~(\ref{eq:classicalCMT}), we see that Eq.~(\ref{eq:quantumCMT}) directly corresponds to Eq.~(\ref{eq:classicalCMT}). This 
yields the association $F_0C^{(0)}/ \sqrt{2} \hbar = f_0 \xi $.
As a result, we find that the value of $F_0$ can be directly estimated from the input driving power 
\be
F_0 = \sqrt{2} \hbar f_0 \xi/A^{(0)} = 2 \sqrt{\hbar \mathcal{P}/Q} \approx 2 \sqrt{\hbar IL^2/Q},
\ee
where in the last line we have estimated the incident driving power from the laser intensity $\mathcal{P} = IL^2$ where $L^2$ is the 
incident area of the cavity; here $I$ is the laser intensity. The incident area can be estimated from the mode volume $V_{\rm mode} = L^3$. For the parameters for mode volume chosen, we arrive at $L \approx 78 \, {\rm nm}$. 

We remark that Eq.~(\ref{eq:cavityhamiltonian}) is the full Rabi Hamiltonian for a cavity mode driven by an external circularly polarized field. We now re-write Eq.~(\ref{eq:cavityhamiltonian}) into the basis of circularly polarized operators $a_L = (a_x - i a_y) /\sqrt{2}$ and $a_R = (a_x + i a_y)/\sqrt{2}$:
\begin{align}
H_{\rm cav} &=  \hbar \omega (a_L^\dag a_L + a^\dag_R a_R) + \frac{F_0}{2 \sqrt{2}} (a_x^\dag + a_x) \left( e^{i \omega_{\rm d} t} +  e^{-i \omega_{\rm d} t} \right) + \frac{F_0}{2 \sqrt{2} i} (a_y^\dag + a_y) \left( e^{i \omega_{\rm d} t} -  e^{-i \omega_{\rm d} t} \right) \nonumber \\
&= \hbar \omega (a_L^\dag a_L + a_R^\dag a_R) + \frac{F_0}{2\sqrt{2}} \left( a_x^\dag - i a_y^\dag + a_x - i a_y \right) e^{i \omega_{\rm d} t} + \frac{F_0}{2 \sqrt{2}} \left( a_x^\dag + i a_y^\dag + a_x + i a_y \right) e^{-i \omega_{\rm d} t} \nonumber \\
&= \hbar \omega (a_L^\dag a_L + a_R^\dag a_R) + \frac{F_0}{2 } \left( a_L e^{i \omega_{\rm d} t} + a_L^\dag e^{-i \omega_{\rm d} t} \right) + \frac{F_0}{2} \left( a_R^\dag e^{i \omega_{\rm d} t} + a_R e^{-i \omega_{\rm d} t} \right). 
\end{align}
This formulation allow us to focus on the component with a single circular polarization by discarding the counter-rotating component in the rotating wave approximation. 

We observe that in the Heisenberg picture, $\partial_t a_{L,R} = i / \hbar [H_{\rm cav} ,a_{L,R}]$, the operator $a_{L,R}$ evolves as $ e^{- i \omega t}$. Similarly, the operator $a_{L,R}^\dag$ evolves as $e^{ i \omega t}$. Thus, the right-handed circularly polarized component $\left( a_R^\dag e^{i \omega_{\rm d} t} + a_R e^{-i \omega_{\rm d} t} \right)$ rotates about twice as fast as $\omega$ when moving to the rotating frame (as employed in the main text). Therefore, within the rotating wave approximation, we can discard the right-handed component and the resulting Hamiltonian is consistent with Eq.~(\ref{eq:driven H}) in the main text. 

\subsection{Multistable $\overline{P}_{\rm tot}$ in a TMD photonic cavity} 

\begin{figure}[h]
    \centering
    \includegraphics[scale=0.4]{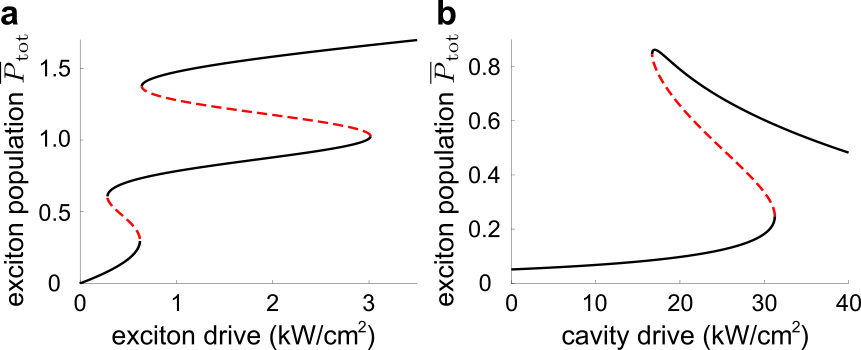}
    \caption{Exciton population $\overline{P}_{\rm tot}$ as a function of exciton drive (panel a) and cavity drive (panel b) exhibiting multiple steady states. The parameters used are the same as those in Fig.~\ref{fig3}b,c of the main text}
    \label{figS}
\end{figure}

In this section, we plot the multi-stable steady state excitonic population as a function of exciton and cavity drives in Fig.~\ref{figS}. This displays how the average steady state $\overline{P}_{\rm tot}$ takes on small values on the order of $\overline{P}_{\rm tot} \sim 1$.
This indicates that the MSE occurs even as approximately one exciton is excited in the entire photonic cavity (corresponding to a low exciton density of order $10^{10}\ {\rm cm}^{-2}$). Further, we find that jumps in the exciton population (see Fig.~\ref{figS}b) can be tuned to be of order unity. 

\begin{thebibliography}{9}

\bibitem{basov}  D. N. Basov, R. D. Averitt, and D. Hsieh, ``Towards properties on demand in quantum materials.'' Nat. Mater. \textbf{16}, 1077-1088 (2017).

\bibitem{stark_morkoc}
A. Mysyrowicz, D. Hulin, A. Antonetti, A. Migus, W.T. Masselink, H. Morkoc, ```Dressed Excitons' in a Multiple-Quantum-Well Structure: Evidence for an Optical Stark Effect with Femtosecond Response Time", Phys. Rev. Lett. {\bf 56}, 2748 (1986) 

\bibitem{stark_chemla}
S. Schmitt-Rink, D.S. Chemla, ``Collective excitations and the dynamical Stark effect in a coherently driven exciton system'', Phys. Rev. Lett. {\bf 57}, 2752 (1986).

\bibitem{stark_townes}
S. H. Autler, C. H. Townes, ``Stark effect in rapidly varying fields", Phys Rev. {\bf 100}, 703-722 (1955).

\bibitem{stark_GaAs}
A. Von Lehmen, D. S. Chemla, J. E. Zucker, and J. P. Heritage, ``Optical Stark effect on excitons in GaAs quantum wells", Opt. Lett. \textbf{11}, 609-611 (1986).

\bibitem{Combescot}
M. Combescot, ``Polariton versus optical stark effect: An old concept revisited", Solid State Communications \textbf{74}, 291 (1990).

\bibitem{Hayat}
A. Hayat, et al., ``Dynamic Stark Effect in Strongly Coupled Microcavity Exciton Polaritons'', Phys. Rev. Lett. {\bf 109}, 033605 (2012).


\bibitem{stark}
E. J. Sie, J. W. McIver, Y. H. Lee, L. Fu, J. Kong, and N. Gedik, ``Valley-selective optical Stark effect in monolayer WS$_2$", Nat. Mater. \textbf{14}, 290 (2015).

\bibitem{polariton_hopfield}
J.J. Hopfield, ``Theory of the contribution of excitons to the complex dielectric constant of crystals'', Phys. Rev. {\bf 112}, 1555 (1958).

\bibitem{polariton_weisbuch}
C. Weisbuch, M Nishioka, A. Ishikawa, Y. Arakawa, ``Observation of the coupled exciton-photon mode splitting in a semiconductor quantum microcavity'', Phys. Rev. Lett. {\bf 69}, 3314 (1992). 

\bibitem{TMD_polariton}
X. Liu, T. Galfsky, Z. Sun, F, Xia, E. Lin, Y. Lee, S. K\'{e}na-Cohen and V. M. Menon, ``Strong light–matter coupling in two-dimensional atomic crystals", Nat. Photon. \textbf{9}, 30-34 (2015). 

\bibitem{stark_BS}
E. J. Sie, C. H. Lui, Y. H. Lee, L. Fu, J. Kong, and N. Gedik, ``Large, valley-exclusive Bloch-Siegert shift in monolayer WS$_2$", Science \textbf{355}, 1066 (2017).

\bibitem{stark_biexciton}
C.-K. Yong \textit{et al.}, ``Biexcitonic optical Stark effects in monolayer molybdenum diselenide", Nat. Phys. \textbf{14}, 1092-1096 (2018).

\bibitem{PhC1}
J. Zhou, J. Zheng, Z. Fang, P. Xu, and A. Majumdar, ``Ultra-low mode volume on-substrate silicon nanobeam cavity", Opt. Express \textbf{27}, 30692 (2019).

\bibitem{PhC2}
R. Miura, S. Imamura, R. Ohta, A. Ishii, X. Liu, T. Shimada, S. Iwamoto, Y. Arakawa, and Y. K. Kato, ``Ultralow mode-volume photonic crystal nanobeam cavities for high-efficiency coupling to individual carbon nanotube emitters", Nat. Commun. \textbf{5}, 5580 (2014).

\bibitem{PhC3}
T. Asano, Y. Ochi, Y. Takahashi, K. Kishimoto, and S. Noda, ``Photonic crystal nanocavity with a Q factor exceeding eleven million", Opt. Express \textbf{25}, 1769 (2017). 

\bibitem{PhC4}
X. Yang, A. Ishikawa, X. Yin, and X. Zhang, ``Hybrid Photonic-Plasmonic Crystal Nanocavities", ACS Nano \textbf{5}, 2831 (2011).

\bibitem{Rudner}
Z. Maizelis, M. Rudner, and M. I. Dykman, ``Vibration multistability and quantum switching for dispersive coupling", Phys. Rev. B \textbf{89}, 155439 (2014). 

\bibitem{TMD_exciton_review}
G. Wang, A. Chernikov, M. M. Glazov, T. F. Heinz, X. Marie, T. Amand, and B. Urbaszek, ``Colloquium: Excitons in atomically thin transition metal dichalcogenides", Rev. Mod. Phys. \textbf{90}, 021001 (2018). 

\bibitem{TMD1}
A. M. Jones \textit{et al.}, ``Optical generation of excitonic valley coherence in monolayer WSe$_2$", Nat. Nanotechnol. \textbf{8}, 634 (2013).

\bibitem{TMD2}
M. M. Glazov, T. Amand, X. Marie, D. Lagarde, L. Bouet, and B. Urbaszek, ``Exciton fine structure and spin decoherence in monolayers of transition metal dichalcogenides", Phys. Rev. B \textbf{89}, 201302 (2014). 

\bibitem{TMD3}
H. Yu, G.-B. Liu, P. Gong, X. Xu, and W. Yao, ``Dirac cones and Dirac saddle points of bright excitons in monolayer transition metal dichalcogenides", Nat. Commun. \textbf{5}, 3876 (2014).

\bibitem{TMD_valley}
J. Kim, X. Hong, C. Jin, S.-F. Shi, C.-Y. S. Chang, M.-H. Chiu, L.-J. Li and F. Wang, ``Ultrafast generation of pseudo-magnetic field for valley excitons in WSe$_2$ monolayers", Science \textbf{346} 1205 (2014).

\bibitem{intervalley_scatter}
G. Kioseoglou, A. T. Hanbicki, M. Currie, A. L. Friedman, and B. T. Jonker, ``Optical polarization and intervalley scattering in single layers of MoS$_2$ and MoSe$_2$", Sci. Rep. \textbf{6}, 25041 (2016).

\bibitem{Rana}
H. Wang, C. Zhang, W. Chan,  C. Manolatou, S. Tiwari, and F. Rana, ``Radiative lifetimes of excitons and trions in monolayers of the metal dichalcogenide 
MoS$_2$", Phys. Rev. B \textbf{93}, 045407 (2016). 

\bibitem{Lee}
Y. C. Lee, and P. S. Lee, ``Coherent radiation from thin films", Phys. Rev. B \textbf{10}, 344 (1974).

\bibitem{Yamamoto}
G. Bj\"{o}rk, S. Pau, J. Jacobson, and Y. Yamamoto, ``Wannier exciton superradiance in a quantum-well microcavity", Phys. Rev. B \textbf{50}, 17336 (1994). 

\bibitem{exciton_lifetime}
H. H. Fang, \textit{et al.}, ``Control of the Exciton Radiative Lifetime in van der Waals Heterostructures", Phys. Rev. Lett. \textbf{123}, 067401 (2019)

\bibitem{WSe2_lifetime}
G. Moody \textit{et al.}, ``Intrinsic homogeneous linewidth and broadening mechanisms of excitons in monolayer transition metal dichalcogenides", Nat. Commun. \textbf{6}, 8315 (2015).

\bibitem{exciton_lifetime_2ps}
C. Robert \textit{et al.}, ``Exciton radiative lifetime in transition metal dichalcogenide monolayers", Phys. Rev. B \textbf{93}, 205423 (2016).

\bibitem{WS2_lifetime}
G. Plechinger, P. Nagler, A. Arora, R. Schmidt, A. Chernikov, A. G. del \'{A}guila, P. C. M. Christianen, R. Bratschitsch, C. Sch\"{u}ller, and T. Korn, ``Trion fine structure and coupled spin–valley dynamics in monolayer tungsten disulfide", Nat. Commun. \textbf{7}, 12715 (2016).

\bibitem{MoS2_hBN}
F. Cadiz \textit{et al.}, ``Excitonic Linewidth Approaching the Homogeneous Limit in MoS$_2$-Based van der Waals Heterostructures", Phys. Rev. X \textbf{7}, 021026 (2017).

\bibitem{polariton_bistability}
A. Baas, J. Ph. Karr, H. Eleuch, and E. Giacobino, ``Optical bistability in semiconductor microcavities", Phys. Rev. A \textbf{69}, 023809 (2004).

\bibitem{bistability_phototherm1}
G. Scuri \textit{et al.}, ``Large Excitonic Reflectivity of Monolayer MoSe$_2$ Encapsulated in Hexagonal Boron Nitride", Phys. Rev. Lett. \textbf{120}, 037402 (2018). 

\bibitem{bonifacio} R. Bonifacio, and L. A. Lugiato, ``Optical bistability and cooperative effects in resonance fluorescence'', Phys. Rev. A. {\bf 18} 1129 (1978).  

\bibitem{hassan1978} S. S. Hassan, P. D. Drummond, and D. F. Walls, ``Dispersive optical bistability in a ring caivty'', Opt. Commun. {\bf 27} 480 (1979).

\bibitem{gibbs} H. M. Gibbs, S. L. McCall, T. N. C. Venkatesan, A. C. Gossard, A. Passner, W. Wiegmann, ``Optical bistability in semiconductors'', Appl. Phys. Lett. {\bf 35}, 451 (1979). 

\bibitem{bowden}
C. M. Bowden, C. C. Sung, ``First- and second-order phase transitions in the Dicke model: Relation to optical bistability'', Phys. Rev. A {\bf 19}, 2392 (1979). 

\bibitem{hermann} 
J. A. Hermann, H. J. Carmichael, ``Stark effect in dispersive optical bistability'', Opt. Letters {\bf 7}, 207 (1982). 

\bibitem{Ginossar2012}
  E. Ginossar, L. S. Bishop, and S. M. Girvin, ``Nonlinear oscillators and high fidelity qubit state measurement in circuit quantum electrodynamics,'' in {\it Fluctuating Nonlinear Oscillators. From nanomechanics to quantum superconducting circuits}, Oxford University Press (2012).

\bibitem{Joannopoulos}
J. D. Joannopoulos, S. G. Johnson, J. N. Winn, and R. D. Meade, \textit{Photonic Crystals: Molding the Flow of Light} (Princeton University Press, 2008).

\end{thebibliography}
\end{document}